\documentclass[envcountsect,envcountsame,runningheads,a4paper]{llncs}

\usepackage{main}
\def\BibTeX{{\rm B\kern-.05em{\sc i\kern-.025em b}\kern-.08em
    T\kern-.1667em\lower.7ex\hbox{E}\kern-.125emX}}
\usepackage{orcidlink}
\hypersetup{pdfborder={0 0 0}}
\usepackage[subtle,tracking=normal]{savetrees}
\newcommand{\cle}{\color{red}}%

\newcommand{\myline}[1]{on line \texttt{#1}}
\newcommand{\mylines}[2]{on lines \texttt{#1-#2}}

\usepackage{graphbox} 

\hypersetup{
    colorlinks,
    linkcolor={red!50!black},
    citecolor={blue!50!black},
    urlcolor={blue!80!black}
}

\newcommand{\ndlimit}{$^\dagger$}
\newcommand{\coderepo}{https://github.com/RandallYe/Animation_of_Security_Protocolss/blob/01341eb3b0e404d86598eef79f916a8eb2940267}
\newcommand{\codereponspkmsg}{\coderepo/NSPK/NSPK3/NSPK3_message.thy}
\newcommand{\codereponspk}{\coderepo/NSPK/NSPK3/NSPK3.thy}
\newcommand{\codereponspkmsgseven}{\coderepo/NSPK/NSPK7/NSPK7_message.thy}

\newcommand{\coderepodh}{\coderepo/Diffie_Hellman}
\newcommand{\coderepodhmsg}{\coderepodh/DH_message.thy}

\AtBeginEnvironment{alltt}{\setlength{\topsep}{0pt}}

\begin{document}

\title{User-Guided Verification of Security Protocols via Sound Animation}


\author{Kangfeng Ye \and Roberto Metere \and Poonam Yadav}
\institute{University of York, York, UK \\ \email{\{kangfeng.ye,roberto.metere,poonam.yadav%
\}@york.ac.uk}}

\authorrunning{Ye et al.}

\maketitle

\begin{abstract}
Current formal verification of security protocols relies on specialized researchers and complex tools, inaccessible to protocol designers who informally evaluate their work with emulators.
This paper addresses this gap by embedding symbolic analysis into the design process.
Our approach implements the Dolev-Yao attack model using a variant of CSP based on Interaction Trees (ITrees) to compile protocols into animators -- executable programs that designers can use for debugging and inspection.
To guarantee the soundness of our compilation, we mechanised our approach in the theorem prover Isabelle/HOL.
As traditionally done with symbolic tools, we refer to the Diffie-Hellman key exchange and the Needham-Schroeder public-key protocol (and Lowe's patched variant).
We demonstrate how our animator can easily reveal the mechanics of attacks and verify corrections.
This work facilitates security integration at the design level and supports further security property analysis and software-engineered integrations.

\textbf{Keywords}: Interaction Trees, CSP, Sound Animation, Formal Verification, Code Generation, Security Protocols
\end{abstract}

\section{Introduction}

The application of formal verification to security protocols has brought clear benefits and provable guarantees of several security properties.
Examples are: 
{\em authentication}, the process of verifying a claimed identity of a user, device, or other entity in a computer system~%
\cite{Basin2018};
{\em handover}, the passing on of responsibilities of a particular post by the outgoing guard to the incoming guard~\cite{Peltonen2021};
{\em privacy} of data and devices against outside threats, spyware, and subversion~\cite{Arapinis2012};
{\em access control}, determining who is allowed to access certain data, applications, and resources, and in what circumstances~\cite{Akon2023}.
The above properties increased the global trust towards popularly adopted protocols, as TLS~\cite{Bhargavan2017,Basin2022}.

A range of tools have been developed or applied to automatically analyse the security properties of protocols, such as FDR~\cite{GibsonRobinson2014}, Isabelle~\cite{Nipkow2002}, Maude-NPA~\cite{Escobar2006}, AVISPA~\cite{Armando2005}, ProVerif~\cite{Blanchet2016}, Tamarin-prover~\cite{Basin2017}.
They have demonstrated useful tools for the discovery of attacks to otherwise unknown vulnerabilities~\cite{Lowe1996,Arapinis2012,
Akon2023}, for the identification of missing or weak assumptions~\cite{Basin2018}, for the proposal of fixes or improvements to protocols~\cite{Lowe1996,Basin2018,Miller2022}, and for the guarantee of correctness~\cite{Peltonen2021}.
These works were carried out by formal verification and security researchers with solid knowledge of formal specification and verification and experience with particular verification tools.
So, the approaches used in these works are \emph{not} obviously accessible by other users, such as security protocol designers.
The general procedure for the application of these approaches starts with a complete security protocol, followed by protocol modelling in a formal specification language, verification using a tool, and finally concludes with a verification report (problems, fixes, suggestions, etc) to be sent to the protocol designers or standard organisers.
This procedure is usually \emph{non-iterative} or \emph{each iteration takes a very long time} because designers do not actively participate in this verification loop.

Several studies~\cite{Kazmierczak1998,Boichut2007,Dutle2015,Miller2022} have investigated a formal and accessible technique, \emph{animation}, to general designers or engineers.
Here, an animation of a formal specification model is an executable computer program implemented in programming languages such as C++, Java, or Haskell.
It provides user interfaces 
to allow users to interact with the encoded model.
This allows users to inspect the model's behaviour by interactively choosing what the model is allowed to do and observing its response. 
A significant problem with the above-mentioned studies is that the animation is {\em not} guaranteed to be sound, that is, they provide no guarantees that all the traces of an animation unequivocally correspond to traces of the original specification.
Soundness becomes an essential aspect if we apply animations to the security of protocols, as one needs to prove that a counterexample or an attack found in the animation must reflect a problem in the original protocol specification.

Our main idea to provide the required soundness is to use Interaction Trees~\cite{Xia2019} (ITrees) where a formal specification can be associated with an abstract and also executable denotational semantics~\cite{Xia2022}.
Using ITrees, Foster et al.~\cite{Foster2021} gave semantics to a version of CSP process algebra~\cite{Hoare1985,Roscoe2011}, called ITree-based CSP, and mechanised it in Isabelle/HOL~\cite{Nipkow2002} to allow automatic generation of Haskell code from a CSP model for animation.
This has been applied in~\cite{Ye2024} to animate control software in robotics
 and demonstrate functional correctness through manual interaction.
The animation is thus sound thanks to ITree's executable semantics and Isabelle's code generator~\cite{Haftmann2010}, which translates executable ITree definitions in the source HOL logic to target functional languages (such as Haskell).
The translation preserves semantic correctness using higher-order rewrite systems~\cite{Mayr1998}.
We apply ITree-based CSP to go beyond functional correctness to obtain sound animation for the security properties of protocols and extend the animator to automatically check reachability and feasibility.
We emphasise that interactivity and automation not only provide verified security insights at the early stages of the protocol design, but they do so while releasing the burden of having to know additional formal languages.
This is a significant step toward the accessibility required by designers and other wider stakeholders.

\begin{figure}[t]
    \begin{center}
        \includegraphics[width=0.85\textwidth]{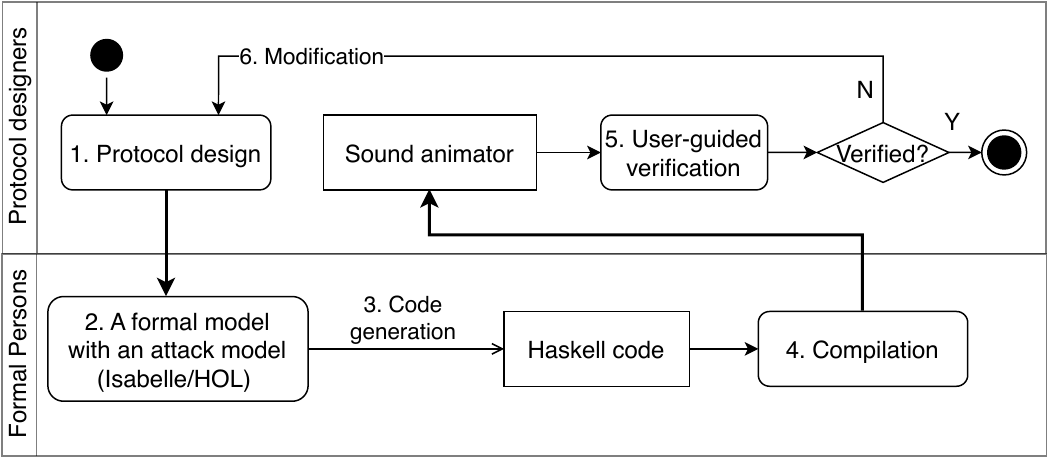}
    \end{center}
    \vspace*{-1em}
    \caption{User-guided verification workflow via sound animation.}
    \vspace*{-2em}
    \label{fig:workflow}
\end{figure}

One workflow that our approach supports is {\em interactive}, as shown in \Cref{fig:workflow}, where protocol designers can use the animator to carry out user-guided verification.
We call this workflow \emph{lightweight verification} because the animation is sound but may not be exhaustive.
Through our animator, the designer can
\begin{enumerate*}[label=(\arabic*),]
    \item manually explore the protocol by interacting with the animator, 
    \item automatically explore the exhaustive event space up to specified steps,
    \item automatically check the reachability of a set of particular events, or
    \item test the feasibility of a specified sequence of events or trace.
\end{enumerate*}
In this way, the animator is indeed a \emph{verified model checker}.

Our technical contributions are summarised as follows: %
\begin{enumerate*}[label=(\arabic*),]
\item we propose a framework to model security protocols 
using ITree-based CSP, which is suitable not only for animation presented in this paper but also for theorem proving in the future;
\item we extend the animator from manual-only exploration to automatic reachability check and feasibility check;
\item we model two security protocols, the Needham-Schroeder public key protocol (NSPK)~\cite{Needham1978} and Diffie–Hellman key exchange protocol (DH)~\cite{Diffie1976}; and %
\item we generate animators and use them to demonstrate attacks and verify patched versions.
\end{enumerate*}
All definitions in this paper are mechanised and show an icon (\isalogo) that links to the corresponding repository artefacts\footnote{We assume basic knowledge of Isabelle/HOL from interested readers to understand these definitions and theorems.}.

The remainder of this paper is organised as follows.
We review related work in \Cref{sec:relwork}, introduce ITrees and ITree-based CSP in \Cref{sec:itree_csp}, and then model and animate the Needham-Schroeder and the Diffie–Hellman protocols in \Cref{sec:nspkp,sec:dhkep} respectively.
Finally, in~\Cref{sec:concl}, we conclude and discuss future work.

\section{Related work}
\label{sec:relwork}
NSPK~\cite{Needham1978} is an important yet simple protocol aiming to establish secure communication between two parties over an insecure network with the help of a trusted third party who securely distributes pre-registered public keys.
Lowe used CSP and the model checker FDR to find a man-in-the-middle attack
 in the protocol (NSPK) and proposed a fix~\cite{Lowe1996} (NSLPK). 
Then, the two versions of the protocol became widely studied examples in the formal verification of security protocols by many researchers using different tools. Paulson~\cite{Paulson1997} used the inductive approach
 in Isabelle/HOL. 
Schneider~\cite{Schneider1997} used CSP hand proofs to analyse the protocol and considered more about cryptographic equations.
Cremers et al.~\cite{Cremers2005} developed operational semantics for security protocols and analysed both versions. Blanchet et al.~\cite{Blanchet2023} used ProVerif, and Meier~\cite{Meier2013a} used Tamarin-prover to analyse the protocols.

DH~\cite{Diffie1976,Merkle1978} is another important protocol on which many authenticated key agreement protocols are based
. DH protocol does not include authentication, making it vulnerable to man-in-the-middle attacks. It is usually combined with other authentication protocols, such as DSA or RSA. 
ProVerif has been applied to verify various DH-based protocols~\cite{Kuesters2009}, and Tamarin is also applied~\cite{Basin2017}.

All these studies used model checking or theorem proving for verification and were conducted by researchers. Instead, in our approach, protocol designers (indeed, anyone after a short training) can verify the protocols using animation.

Kazmierczak et al.~\cite{Kazmierczak1998} developed an animator for Z specifications and used it to explore and test models. They regard animation as a lightweight formal method. Dutle et al.~\cite{Dutle2015} proposed an approach to manually translate formally verified models
 to Java code, which allows the animation of the formal specifications.  
Miller et al.~\cite{Miller2022} implemented an emulator, written in C++, to experiment with 5G authenticated key establishment procedures. 
Boichut et al.'s SPAN~\cite{Boichut2007}, written in OCaml and Tcl/Tk, is an animation tool for the specification language 
used in AVISPA. It can find the attack in NSPK.
 However, the soundness guarantee is the most significant difference of our work from SPAN and the other.
Additionally, our approach is more flexible and accessible to extend (for example, to support different attack models) and maintained because animators are automatically generated from CSP specifications. 

Our work is based on~\cite{Foster2021,Ye2024}, and applies to the modelling and animation of security protocols. Additionally, we extended the animator command line interface from manual-only exploration to combined manual, random, and automatic exploration with reachability and feasibility checks.

\section{ITree-based CSP}
\label{sec:itree_csp}

This section briefly introduces interaction trees and the ITree-based CSP processes and operators, with their semantics omitted here for brevity. A complete account of the semantics is in \cite{Foster2021,Ye2024}.

Interaction trees (ITrees)~\cite{Xia2019} are a data structure for modelling reactive systems interacting with their environment through events. They are potentially \emph{infinite} and defined as coinductive trees
in Isabelle/HOL.
\newcommand{\itreedef}{\isalink{https://github.com/isabelle-utp/interaction-trees/blob/418b37554f808828610f10b40c051a562fe0c716/Interaction_Trees.thy\#L22}}
\begin{alltt}
\isakwmaj{codatatype} (\textquotesingle{e}, \textquotesingle{r}) itree = \(\itreedef\) 
  Ret \textquotesingle{r} | Sil "(\textquotesingle{e}, \textquotesingle{r}) itree" | Vis "\textquotesingle{e} \(\pfun\) (\textquotesingle{e}, \textquotesingle{r}) itree"
\end{alltt}
\noindent ITrees are parameterised over two types: \texttt{\textquotesingle e} for events ($E$), and \texttt{\textquotesingle r} for return values or states ($R$) . Three possible interactions are provided: 
\begin{inparaenum}[(1)]
\item $\Ret~x$: termination with a value $x$ of type $R$ returned, denoted as $\tickv{x}$;
	\item $\Sil~P$: an internal silent event, denoted as $\tau P$ for a successor ITree $P$; or
	\item $\Vis~F$: a choice among several visible events represented by a partial function $F$ of type $E \pfun (E,R) \cspkey{itree}$.
\end{inparaenum}

In~\cite{Foster2021,Ye2024}, deterministic CSP processes are given executable semantics using ITrees. Determinism is inherent in ITrees since we use a partial function to model events and their continuations. 
The CSP operators, therefore, cannot introduce nondeterminism.
We summarise the processes and operators in both traditional (that is, the CSP presented in most literature like~\cite{Hoare1985,
Roscoe2011}) and ITree-based CSP in the first and second columns in \Cref{table:csp_operators} where \ndlimit-marked operators block the events that will lead to nondeterminism in their counterparts in standard CSP, for the sake of determinism. The semantics of $\extchoice$, $\hide$, and $\interrupt$ operators follows the maximal progress assumption~\cite{Foster2021,Ye2024} that $\tau$ has the highest priority, $\Ret$ has higher priority, and then $\Vis$ events have lower priority. 

\begin{table}[t]
  \setlength\extrarowheight{0pt} 
  \caption{\label{table:csp_operators} {Processes and operators in standard CSP and ITree-based CSP
.}}
  \resizebox{\textwidth}{!}{
  \begin{tabularx}{\textwidth}{p{2.0cm}|p{2.1cm}|X}
    \hline
Standard  &
ITree-based &
Description 
\\
\hline
$\tau$ &
$\tau$ &
Invisible event.
\\
$Skip$ or $\tick$ &
\cspkey{skip} (or \Ret ()) &
Skip: terminate immediately, and return a unit type $()$.
\\ 
&
$\Ret~v$ &
Return: terminate immediately and return value $v$.
\\ 
$Stop$ &
\cspkey{stop} &
Deadlock: refuse any interaction without a state change. 
\\ 
$c \then P$ &
$\isakwmin{do} \{ \cspkey{outp}\;c\;(); P\}$ &
Prefix: synchronise on channel $c$, then behave like $P$. 
\\
$c!v \then P$ &
$\isakwmin{do} \{ \cspkey{outp}\;c\;v ; P\}$ &
Output:
synchronise on channel $c$ with value $v$, and then behave like $P$.
\\
$c?x \then P(x)$ &
$\isakwmin{do} \{ x \leftarrow \cspkey{inp}\ c\ \mathbb{U}; P(x)\}$&
Input: accept an input of any value (of type $T$) on channel $c$,  
record it in $x$, and then behave like $P(x)$. 
\\
$c?x:S \then P(x)$ &
$\isakwmin{do} \{ x \leftarrow \cspkey{inp}\ c\ S; P(x)\}$&
Restricted input: 
It is similar to input but only accepts the values from set $S$.
\\
$b \guard P$ &
$\isakwmin{do} \{ \cspkey{guard}\;b; P\}$ &
Guarded process: 
behave like $P$ if $b$ is true or deadlock
\\
$P \extchoice Q$ &
$P \extchoice Q$\ndlimit &
External choice:
offer the environment a choice of the first events of $P$ and $Q$, then behave accordingly. 
\\
$P \intchoice Q$ &
&
Internal choice: 
nondeterministic choice between $P$ and $Q$ without offering the environment a choice.
\\
$P ; Q$ &
$P \fatsemi Q$ &
Sequential composition:
behave like $P$ initially, and behave like $Q$ if $P$ terminates. 
\\
$P \interrupt Q$ &
$P \interrupt Q$\ndlimit &
Interrupt:
behave like $P$, but offer the environment a choice of the initial events of $Q$ at any time until $P$ terminates. If one of these events is performed, $Q$ takes over and behaves accordingly.
\\
$ P \exception{E} Q$ &
$\except{P}{E}{Q}$ &
Exception:
behave like $P$ until $P$ performs an event from set $E$, at that point, then behave like $Q$. 
\\
$P \parallel_{E} Q$ &
$P \parallel_E Q$\ndlimit &
Parallel composition:
$P$ and $Q$ run simultaneously, synchronise on the events in $E$, progress independently on the events not in $E$, and terminate if both terminate. 
\\
$P \interleave Q$ &
$P \interleave Q$\ndlimit &
Interleave:
equal to $P \parallel_{\emptyset} Q$ where $P$ and $Q$ always progress independently on any event.
\\
$ P \hide E$ &
$ P \hide E$\ndlimit &
Hiding:
behave like $P$ except that the events from $E$ become internal.
\\
$ \rename{P}{c \becomes d}$ &
$\rename{P}{\rho}$\ndlimit &
Renaming:
rename the event $c$ in $P$ to $d$, or a relation $\rho$ in ITrees.
\\
Recursive functions, omitted here
&
$\cspkey{iterate}\;b~P~s$ & Tail-recursive iteration: 
continue to execute $P$ while the condition $b~s$ holds and otherwise terminates and returns the current state $s$. 
\\
$I(s) = P; I(s')$ &
$\cspkey{loop}\;P$ &
Infinite loop, equal to $\cspkey{iterate}~b~P~s$ where $b~s$ is always true.
\\
    \hline
  \end{tabularx}}
\end{table}
In addition to nondeterminism, there are some other differences between the two versions of CSP. First, standard CSP processes are stateless, and values are passed between processes only through communication or parameters. In contrast, ITree-based processes are stateful, so they could return values, such as $\tickv{v}$, and then pass in variables in sequential composition. Second, sequential composition $P \fatsemi Q$ in ITrees is defined through a monadic bind operator~\cite{Foster2021}. 
With the monadic \isakwmin{do} notation, we can write a sequential composition like $\isakwmin{do} \{ x \leftarrow \cspkey{inp}~c~S; P(x)\}$ denoting the process accepts input values from set $S$ on channel $c$ and records the value in the variable $x$, then passes the value of $x$ to $P(x)$. Finally, ITree-based CSP provides the tail-recursive iteration $\cspkey{iterate}$ and its special infinite \cspkey{loop}. In standard CSP, this is implemented using recursion.

\section{Needham-Schroeder Public Key Protocol}
\label{sec:nspkp}
\begin{figure}[ht!]
    \subfloat{
  \begin{minipage}[b]{0.1\linewidth}
  \end{minipage}
}
\subfloat{
  \begin{minipage}[b]{0.4\linewidth}
\begin{center}
    \begin{enumerate}[label={{{\arabic*}.}}]
        \item {\cle $ A \to S: (A, B) $}
        \item {\cle $S \to A: \digsig{(\pk{k}{B}, B)}{\sk{k}{S}}$} 
        \item $ A \to B: \aenc{(na, A)}{\pk{k}{B}} $
        \item {\cle $ B \to S: (B, A)$} 
        \item {\cle $ S \to B: \digsig{(\pk{k}{A}, A)}{\sk{k}{S}} $}
        \item $ B \to A: \aenc{(na, nb)}{\pk{k}{A}} $
        \item $ A \to B: \aenc{nb}{\pk{k}{B}} $
\end{enumerate}
\end{center}
  \end{minipage}
}
\subfloat{
  \begin{minipage}[b]{0.5\linewidth}
    \begin{center}
        \includegraphics[width=0.8\textwidth]{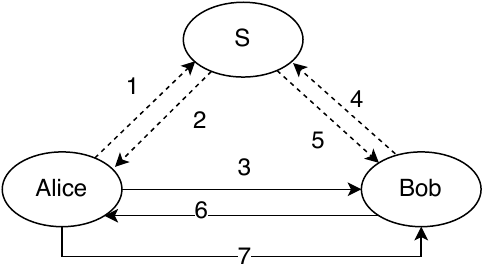}
    \end{center}
  \end{minipage}
}
    \caption{Needham-Schroeder public key protocol}
    \label{fig:nspkp}
\end{figure}

NSPK~\cite{Needham1978} is used to establish mutual authentication between two parties (such as Alice and Bob) over an insecure or public network using asymmetric encryption. Alice and Bob do not know each other's public keys and use a trusted server (S) to distribute them on request. The protocol is illustrated in \Cref{fig:nspkp} where a sequence of seven messages is shown on the left, and the interaction between A (Alice), B (Bob), and S is displayed on the right.

In this protocol, A and B are assumed to know the public key of $S$, which is used to verify messages from S that are digitally signed using the private key of S. 
Among the seven messages, four of them are used by A (or B) to request the public key of B (or A) from S in Message 1 (or 4) and by S to return the B's (or A's) public key $\pk{k}{B}$ (or $\pk{k}{A}$) to A (or B) in Message 2 (or Message 5), which is signed using a private key ${\sk{k}{S}}$ of S. These messages are for \emph{public key retrieve}. In the other three messages, A sends B Message 3 composed of her nonce ($na$) and identity (A) and encrypted using B's public key $\pk{k}{B}$. B replies A Message 6 composed of $na$ and his nonce ($nb$) and encrypted using A's public key $\pk{k}{A}$, and finally, A confirms to B that she knows $nb$ in Message 7. The other three messages are for \emph{authentication}.

This protocol's security goals include the secrecy or confidentiality of Alice's and Bob's nonces (that is, $na$ and $nb$) along with mutual authentication.
This protocol, named NSPK7 here, is well-known to be vulnerable to a man-in-the-middle attack~\cite{Lowe1996}.
Most researchers studied a simplified 3-message version, NSPK3, where the public key retrieve messages are omitted because communications with the trusted server are assumed to be secure (through a secure channel).
We also use model NSPK3 and show how our animation can detect the attack automatically.
Before showing that, we need to introduce the network and attack models that characterise the Dolev-Yao analysis.

\begin{figure}[ht!]
    \begin{center}
        \includegraphics[width=0.5\textwidth]{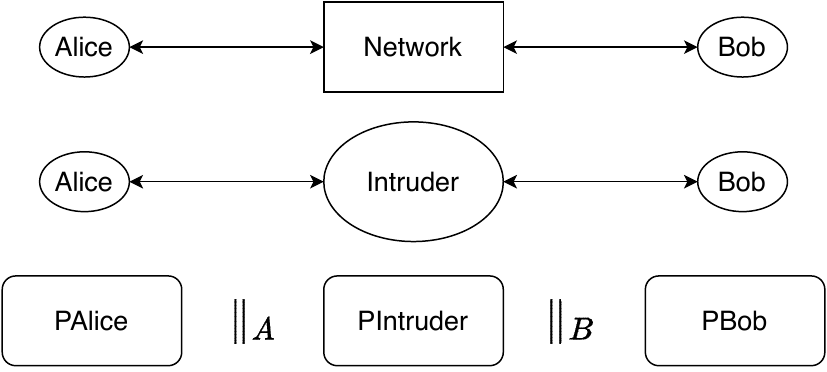}
    \end{center}
    \caption{Network model.}
    \label{fig:network}
\end{figure}

\paragraph{Network model.}
Alice and Bob communicate through an insecure network (at the top of \Cref{fig:network}), which is under the attacker's control.
Thus, Alice and Bob send and receive messages to and from the attacker (the diagram in the middle). In CSP, we use parallel composition to model communication (the diagram at the bottom).

\paragraph{Attack model.}
We consider the Dolev-Yao model~\cite{Dolev1983} for symbolical verification of security protocols where the attacker or intruder controls the entire network and can intercept, delete, modify, delay, inject, and build new messages from the current knowledge but is limited by the (perfect black-box) cryptography that the intruder cannot decrypt a message without knowing its key and forge a signature.
As such, the intruder can augment their knowledge offline by applying inference rules to any pieces of knowledge he already has. Such rules are defined in \Cref{table:inference} where $K$ denotes its current knowledge. Rules are divided into two groups: \emph{break-down} ($\inferd$) rules, what messages can be derived from $K$ using member, unpairing, decryption, and digital signature verification rules, and \emph{build-up} ($\inferu$) rules, what messages can be built up from $K$ using member, pairing, and encryption, and digital sign rules.

\begin{table}[ht]
    \caption{Intruder message inference rules.}
    \label{table:inference}\centering
\bgroup
\def\arraystretch{1.0}
\setlength\tabcolsep{.5mm}
    \begin{tabular}{@{}l cl cl cl cl @{}}
        \toprule
Name & \phantom{a} & Premise & \phantom{a}  & Break down & \phantom{a} & Build up & \phantom{a} & \\
        \midrule
Member & & $m \in K$ && $K \vdash_{\Downarrow} m$ && $K \vdash_{\Uparrow} m$ && \\
Pairing &&$m \in K \land m' \in K$ && && $ K \vdash_{\Uparrow} (m,m')$ && \\
Unpairing &&$(m,m') \in K$ && $K \vdash_{\Downarrow} m$ and $K \vdash_{\Downarrow} m$ && &&\\
Encryption/Sign && $m \in K \land k \in K $ && && $K \vdash_{\Uparrow} \{m\}_k$ && \\
Decryption/Verify && $\{m\}_k \in K \land k^{-1} \in K$ && $K \vdash_{\Downarrow} m$ && && \\
        \bottomrule
\end{tabular}
\egroup
\end{table}
We note that the encryption and decryption inference rules suit asymmetric, symmetric, and digital signatures. Symmetric encryption uses a single key for encryption and decryption (its inverse $k^{-1}=k$). Asymmetric encryption uses a public key for encryption and a private key for decryption (so ${k^{-1}}^{-1}=k$). A digital signature uses a private key to sign a message and a public key to verify authenticity.

\paragraph{Modelling of NSPK3.}
Below, we define a variety of data types used in this protocol.
\newcommand{\dagent}{\isalink{\codereponspkmsg\#L11}}
\begin{alltt} 
\isakwmaj{datatype} dagent = Alice | Bob | Intruder \dagent
\isakwmaj{datatype} dnonce = N dagent
\isakwmaj{datatype} dpkey = PK dagent 
\isakwmaj{datatype} dskey = SK dagent
\isakwmaj{datatype} dkey = Kp dpkey | Ks dskey
\isakwmaj{datatype} dmsg = MAg (ma:dagent) | MNon (mn:dnonce) 
  | MKp (mkp:dpkey) | MKs (mks:dskey) 
  | MCmp (mc1:dmsg) (mc2:dmsg) | MEnc (mem:dmsg) (mek:dpkey)
\end{alltt}
The enumeration \cspcode{dagent} contains three considered agents. Each agent is associated with a nonce of type \cspcode{dnonce} using a constructor \cspcode{N}, a public key of type \cspcode{dpkey} using a constructor \cspcode{PK}, and a private key of type \cspcode{dskey} using a constructor \cspcode{SK}.  Both public keys and private keys are of a type \cspcode{dkey} using different constructors \cspcode{Kp} and \cspcode{Ks}. All messages for communication on channels are of type \cspcode{dmsg}, including agent's identities of type \cspcode{dagent} (with the constructor \cspcode{MAg}), nonces (with \cspcode{MNon}), public keys (with \cspcode{MKp}), private keys (with \cspcode{MKs}), compound of two messages (with \cspcode{MCmp}), and asymmetric encryption (with \cspcode{MEnc}). We note \cspcode{ma}, \cspcode{mn}, etc. in the definition of \cspcode{dmsg} are functions used to extract corresponding elements from a message of type $\cspcode{dmsg}$. For example, $\cspcode{ma}$ has a type \cspcode{dmsg$\Rightarrow$dagent} which returns the agent identity $\cspcode{Alice}$ encoded in a message like \cspcode{MAg Alice}. We also introduce the symbols $\mcomp{m_1}{m_2}$ and $\aenc{m}{k}$ for the compound \cspcode{MCmp m1 m2} and the encryption \cspcode{MEnc m k}. We also use $\langle m_1, m_2, m_3 \rangle$ for $\mcomp{m_1}{\mcomp{m_2}{m_3}}$. 

Inspired by~\cite{Lowe1996}
, we declare signals to specify and verify properties . 
\newcommand{\dsig}{\isalink{\codereponspkmsg\#L214}}
\begin{alltt}
\isakwmaj{datatype} dsig = StartProt dagent dagent dnonce dnonce \dsig 
  | EndProt dagent dagent dnonce dnonce
  | ClaimSecret (sag:dagent) (sn:dnonce) (sp: "\(\power\) dagent")
\end{alltt}
Using \cspcode{ClaimSecret A na \{B\}}, for example, the agent \cspcode{A} claims its nonce \cspcode{na} only known to \cspcode{B}. We use \cspcode{StartProt} or \cspcode{EndProt A B na nb}  to signal \cspcode{A} starts or finishes a protocol run with \cspcode{B} using their nonces \cspcode{na} and \cspcode{nb}.

\paragraph{Message inferences.} In Isabelle, we define a function \cspcode{break\_sk} for the break-down rules in \Cref{table:inference} to derive a list of messages from a given list and supplied set (\cspcode{sks}) of private keys for decryption, for the sake of implementation in Haskell.
%
\newcommand{\breaksk}{\isalink{\codereponspkmsg\#L247}}
\begin{alltt}
\isakwmaj{fun} break_sk::"dmsg list \(\Rightarrow\) dskey set \(\Rightarrow\) dmsg list" \isakwmin{where}  \breaksk
break_sk [] sks = []  |
break_sk \(\mcomp{m1}{m2}\)#xs sks = remdups 
  (break_sk [\(m1\)] sks @ break_sk [\(m2\)] sks @ break_sk xs sks) |
break_sk \(\aenc{m}{\small (PK k)}\)#xs sks = if SK k \(\in\) sks then 
  insert \(\aenc{m}{\small (PK k)}\) (remdups (break_sk [\(m\)] sks @ break_sk xs sks))
  else insert \(\aenc{m}{\small (PK k)}\) (break_sk xs sks) |
break_sk m#xs sks = insert m (break_sk xs sks)
\end{alltt}
For a compound message (\cspcode{$\mcomp{m1}{m2}$} in the head of a list), the unpairing rule in \Cref{table:inference} is applied: recursively break-down \cspcode{m1}, \cspcode{m2}, and also the tail \cspcode{xs}, concatenate (\cspcode{@}) them together with the duplicates removed by \cspcode{remdups}. 
For asymmetric encryption, the decryption rule is applied depending on whether the corresponding private key (\cspcode{SK k}) is in \cspcode{sks} or not, where \cspcode{insert} is a list function to insert an element into a list if it does exist in the list or ignore it. We apply the member rule for other messages (using the pattern match \cspcode{m$\#$xs}). We also define \cspcode{breakm xs} be \cspcode{break\_sk xs (extract\_skey xs)} where \cspcode{extract\_skey} extracts private keys from the list \cspcode{xs} of messages.

For the build-up rules, the resulting messages could be very large or infinite by, for example, applying the pairing and encryption rules alternately. Dealing with these large sets of messages will take space in memory and computation time. For the sake of animation, we limit the number of times these rules will be applied in the definition \cspcode{build\_n} below where \cspcode{build\_n xs ks nc ne l} is used to build up messages from a given list \cspcode{xs} of messages, a list \cspcode{ks} of public keys, the number \cspcode{nc} of applications of the pairing rule, the number \cspcode{ne} of applications of the encryption rule, and the maximum length \cspcode{l} of built messages. However, this will not impact the correctness of protocols verified with appropriate parameters. For example, for NSPK3 or NSPK7, \cspcode{nc}, \cspcode{ne}, and \cspcode{l} could be 1, 1, and 2.

\newcommand{\buildn}{\isalink{\codereponspkmsg\#L322}}
\begin{alltt}
\isakwmaj{fun}{} build_n ::"dmsg list\(\Rightarrow\)dpkey list\(\Rightarrow\)nat\(\Rightarrow\)nat\(\Rightarrow\)nat\(\Rightarrow\)dmsg list"  \buildn
\isakwmin{where} build_n xs ks 0 0 l = xs |
build_n xs ks (Suc nc) 0 l = 
    (build_n (union (pair2 xs xs l) xs) ks nc 0 l) |
build_n xs ks 0 (Suc ne) l = 
    (build_n (union  (enc_1 xs ks) xs) ks 0 ne l) |
build_n xs ks (Suc nc) (Suc ne) l = (List.union 
    (build_n (union (pair2 xs xs l) xs) ks nc (Suc ne) l)
    (build_n (union (enc_1 xs ks) xs) ks (Suc nc) ne l))
\end{alltt}
The pattern matches on line 2, lines 3 and 4, 5 and 6, and 7-9 above correspond to the member build-up rule, the pairing rule, the encryption rule, and both pairing and encryption rules in \Cref{table:inference}. We define the function \cspcode{pair2 xs ys l} to pair each element \cspcode{x} from \cspcode{xs} with each element \cspcode{y} from \cspcode{ys} to form a list of new compound messages $\mcomp{\cspcode{x}}{\cspcode{y}}$ if \cspcode{x} and \cspcode{y} are not the same, and the length of the new message does exceed \cspcode{l}. The pairing definition on lines 3 and 4 denotes the application of the pairing rule once and then builds up using the pairing rule for one less time (\cspcode{nc}) on the union (on the list) of the first application result (\cspcode{pair2 xs xs l}) and \cspcode{xs}.
The function \cspcode{enc\_1 xs ks} aims to build up messages by encrypting each message in \cspcode{xs} using every public key in \cspcode{ks}. 

\paragraph{Channels.} We define all channels grouped in a type \cspcode{Chan} for communicate below. 
\newcommand{\chan}{\isalink{\codereponspkmsg\#L354}}
\begin{alltt}
\isakwmaj{datatype} Chan = \chan 
  env :: dagent\(\cross\)dagent 
  send, recv, hear, fake :: dagent\(\cross\)dagent\(\cross\)dmsg 
  leak :: dmsg    sig :: dsig   terminate :: unit
\end{alltt}
The channel \cspcode{env} is used for the environment to request an authentication between two agents. The channels \cspcode{send}, \cspcode{recv}, \cspcode{hear}, and \cspcode{fake} are for communication between agents in the form of a source agent, a destination agent, and a message. The \cspcode{leak} channel indicates a secret known by the intruder. The \cspcode{sig} channel signals a particular point (modelled as a signal) reached. The \cspcode{terminate} channel is used to terminate the protocol run.

\paragraph{Processes.}
We model Alice as an initiator using the standard syntax for simplicity.
\newcommand{\initiator}{\isalink{\codereponspk\#L36}}
\begin{align*}
Initiator&(A, na) \defs env!A?B \then sig!(\cspcode{ClaimSecret A na \{B\}}) \then \initiator\\
& send!A!Intruder!\aenc{\mcomp{\cspcode{MNon na}}{\cspcode{Mag A}}}{\cspcode{PK B}} \then \\
& recv!Intruder!A?m:\{nb : dnonce @ \aenc{\mcomp{\cspcode{MNon na}}{\cspcode{MMon nb}}}{\cspcode{PK A}}\} \then \\
& sig!(\cspcode{StartProt A B na nb}) \then 
send!A!Intruder!\aenc{\cspcode{MNon nb}}{\cspcode{PK B}} \then \\ 
& sig!(\cspcode{EndProt A B na nb}) \then terminate \then \cspkey{skip}
\end{align*}
The $Initiator$ has two parameters: $A$ for the agent identity and $na$ for a nonce of $A$. The $Initiator$ first waits for the environment's input on channel \cspcode{env}, where B is the counterpart for the mutual authentication. Then it signals a \cspcode{ClaimSecret} that $na$ of $A$ is only known to $B$; it sends Message 3 in \Cref{fig:nspkp} to the public network (that is, the Intruder) and receives Message 6 from \cspcode{Intruder} using a restricted set of messages; it signals a \cspcode{StartProt} for $A$ to run the protocol with $B$ using \cspcode{na} and \cspcode{nb} where \cspcode{nb} is extracted from \cspcode{m} using \cspcode{mn (mc2 (mem m))}; it sends Message 7 (\cspcode{nb}); then it signals a \cspcode{EndProt} for $A$ to finish the protocol run with $B$ using \cspcode{na} and \cspcode{nb}; and finally the \cspcode{Initiator} terminates.

Similarly, we model Bob as a responder.
\newcommand{\responder}{\isalink{\codereponspk\#L116}}
\begin{align*}
Re&sponder(B, nb) \defs \responder\\ 
& recv!Intruder!B?m:\{na : dnonce; A:dagent @ \aenc{\mcomp{\cspcode{MNon na}}{\cspcode{MAg A}}}{\cspcode{PK B}}\} \then \\
& sig!(\cspcode{ClaimSecret B nb \{A\}}) \then 
 sig!(\cspcode{StartProt B A na nb}) \then \\
& send!B!Intruder!\aenc{\mcomp{\cspcode{MNon na}}{\cspcode{MMon nb}}}{\cspcode{PK A}} \then \\
& recv!Intruder!B?m:\{\aenc{\cspcode{MNon nb}}{\cspcode{PK B}}\} \then \\
& sig!(\cspcode{EndProt B A na nb}) \then terminate \then \cspkey{skip}
\end{align*}
The $na$ and $A$ are extracted from the received message $m$ using \cspcode{ma (mc2 (mem m))} and \cspcode{mn (mc1 (mem m))}.
Given its identity $I$, nonce $ni$, a list $k$ of knowledge, and a list $s$ of secrets, we model Intruder below.
\newcommand{\intruderzero}{\isalink{\codereponspk\#L175}}
\begin{align*}
P&Intruder0(I, ni, k, s) \defs \intruderzero\\
& \left(hear?\_!Intruder?m \then PIntruder0(I, ni, \cspcode{breakm(insert m k)}, s)\right) \extchoice \\
& \left(\Extchoice A:dagent; m:build\_n(k) @ fake!Intruder!A!m \then PIntruder0(I, ni, k, s)\right) \extchoice \\
& \left(\Extchoice m: \cspcode{filter ($\lambda$x. member k x) s} @ leak!m \then PIntruder0(I, ni, k, s)\right) \extchoice \\
& \left(terminate \then \cspcode{skip}\right)
\end{align*}
The process models the following behaviours of the intruder: it can hear a message $m$ from any agent ($\_$) and learn new knowledge (\cspcode{breakm(insert m k)}) from $m$; it can fake all the messages which can be built up from its knowledge; it can leak a message $m$ if $m$ is in both $k$ and $s$ (by \cspcode{filter ($\lambda$x. member k x) s}), that is, a secret known to the intruder; and it can terminate. This process models an active attacker. We can simply remove its capability of faking messages (so no need to build up messages) to model a passive attacker. In this process, if a secret is leaked, it continues to offer the \cspcode{leak!m} event. This is not an incorrect behaviour. But it will result in more traces in the animator later having interleaving \cspcode{leak} events (so it is more challenging to explore the traces using animation). We introduce the processes below to restrict each secret from being leaked only once.

%
\newcommand{\intruderone}{\isalink{\codereponspk\#L213}}
\begin{align*}
    & PLeakOnce(secrects) \defs \Interleave s: secrets@ leak!s \then skip \\
    & {PIntruder1}(I, ni,k,s) \defs \\
    & \ \ \left(PIntruder0(I,ni,k,s)\parallel_{LkEvents(s)} PLeakOnce(s)\right) \except{}{TermEvent}{skip} 
\end{align*}
$PLeakOnce$ is an indexed interleaving and it \cspcode{leak}s each secret $s$ in $secrets$ once. $PIntruder1$ uses the parallel composition to restrict $PIntruder0$ for the \cspcode{leak} event where $LkEvents(s)$ is a set of \cspcode{Leak} events for all secrets $s$. We also use the exception operator to allow the composition to be terminated by a \cspcode{terminate} event.

The initial knowledge of the intruder includes the identities and public keys of all the agents and a private key of the intruder. The secrets of NSPK3 are the nonces $na$ and $nb$ in \Cref{fig:nspkp}.
\newcommand{\initknows}{\isalink{\codereponspk\#L27}}
\newcommand{\allsecrets}{\isalink{\codereponspk\#L25}}
\begin{alltt}
\isakwmaj{definition} InitKnows = \initknows
  AllAgentsLst @ AllPKsLst @ [MNon (N Intruder), MKs (SK Intruder)]
\isakwmaj{definition} AllSecrets = removeAll (MNon (N Intruder)) AllNoncesLst \allsecrets 
\end{alltt}
The behaviour of the intruder $PIntruder$ is defined below with \cspcode{hear} and \cspcode{fake} events renamed to \cspcode{send} and \cspcode{recv} according to a relation $nameMapI$.
\newcommand{\pintruder}{\isalink{\codereponspk\#L227}}
\begin{align*}
    & PIntruder2 \defs PIntruder1~\cspcode{Intruder (N Intruder) InitKnows AllSecrets} \\
    & PIntruder \defs \rename{PIntruder2}{nameMapI}     \pintruder
\end{align*}

In NSPK3, Alice plays the initiator role, and Bob plays the responder role. Their behaviours are the instantiation of \cspcode{Initiator} and \cspcode{Responder}, respectively. 
\begin{alltt}
\isakwmaj{definition} PAlice = Initiator Alice (N Alice)
\isakwmaj{definition} PBob = Responder Bob (N Bob)
\end{alltt}
Finally, the NSPK3 protocol is modelled below where $TermEvent$ contains only the \cspcode{terminate} event, and $ABIEvents$ contains the events for communication between Alice, Bob, and Intruder including the \cspcode{terminate} event.
\newcommand{\pnspk}{\isalink{\codereponspk\#L263}}
\begin{align*}
    {NSPK3} \defs \left({PAlice} \parallel_{TermEvent} {PBob}\right)\parallel_{ABIEvents} {PIntruder} \quad\pnspk
\end{align*}

\paragraph{Animation.}
After NSPK3 is modelled in Isabelle, we can (soundly) generate Haskell code and compile it into an executable program using the \cspcode{animate\_sec}\isaref{{\codereponspk\#L266}} command, which we also implement to animate security protocols using ITrees.

\begin{figure}[t]
\begin{lstlisting}[language=Animation, caption={}, label=]
Starting ITree Animation...
Events:
 (1) Env [Alice] Bob;
 (2) Env [Alice] Intruder;
 (3) Recv [Bob<=Intruder] {<N Intruder, Alice>}_PK Bob;
 (4) Recv [Bob<=Intruder] {<N Intruder, Intruder>}_PK Bob;
[Choose: 1-4]: 1 Env_C (Alice,Bob)
Events:
 (1) Recv [Bob<=Intruder] {<N Intruder, Alice>}_PK Bob;
 (2) Recv [Bob<=Intruder] {<N Intruder, Intruder>}_PK Bob;
 (3) Sig ClaimSecret Alice (N Alice) (Set [ Bob ]);
\end{lstlisting}
\caption{\label{fig:animate-nspk3} First interaction with the NSPK3 animator.}
\vspace{-3ex}
\end{figure}
In \Cref{fig:animate-nspk3}, we show the first interaction of the model with its environment: the lines starting with \lstinline[language=Animation]{Events} are produced by the model or the animator and describe all enabled events, and the lines beginning with \lstinline[language=Animation]{[Choose: 1-n]} represents a user's choice of enabled events from number 1 to n. Initially, there are four enabled events: the first two events \mylines{3}{4} are due to \cspcode{PAlice} to get a request from the environment on which principal to establish a session for Alice, and the other two events \mylines{5}{6} represent the (fake) messages sent by the intruder to Bob. These messages are built up using the intruder's initial knowledge (because no message is heard by now). After one event (the first event in this case) is chosen \myline{7}, then the model accepts the request on the \cspcode{Evn} channel, and three events are enabled to allow users to interact further. This is the default way to explore the protocol manually.

Additionally, the animator provides five extra ways to exhaustively (\cspcode{Auto}) or randomly (\cspcode{Rand}) explore the event space up to specified steps or check the reachability of a set of particular events while monitoring the reachability of other events using both exhaustive (\cspcode{AReach}) or random search (\cspcode{RReach}), or check the feasibility of a specified sequence of events.

We use \cspcode{AReach 15 \%Terminate\%} to search all possible traces for a successful protocol termination, and many traces are found. We use \cspcode{AReach 15 \%Leak N Bob\%} to automatically search the possible violation of the \emph{secrecy} for the nonce of Bob. Only one trace is shown in \Cref{fig:animate-nspk3-violet-secrecy}. This is the man-in-the-middle attack~\cite{Lowe1996}. Similarly, we checked the authenticity of Alice and Bob and found that Alice's authenticity was not violated, but Bob's authenticity was violated in six traces.
\begin{figure}[t]
\begin{lstlisting}[language=Animation, caption={}, label=]
AReach 15,  %Leak N Bob%
Reachability by Auto: 15
  Events for reachability check: ["Leak N Bob"]
  Events for monitoring: []
*** These events ["Leak N Bob"] are reached! ***
Trace: [Env [Alice] Intruder, Sig ClaimSecret Alice (N Alice) (Set [ Intruder ]), Send [Alice=>Intruder] {<N Alice, Alice>}_PK Intruder, Recv [Bob<=Intruder] {<N Alice, Alice>}_PK Bob, Sig ClaimSecret Bob (N Bob) (Set [ Alice ]), Sig StartProt Bob Alice (N Alice) (N Bob), Send [Bob=>Intruder] {<N Alice, N Bob>}_PK Alice, Recv [Alice<=Intruder] {<N Alice, N Bob>}_PK Alice, Sig StartProt Alice Intruder (N Alice) (N Bob), Send [Alice=>Intruder] {N Bob}_PK Intruder, ]
\end{lstlisting}
\caption{\label{fig:animate-nspk3-violet-secrecy} Violation of secrecy for the nonce of Bob.}
\vspace{-3ex}
\end{figure}

Our animation can also check whether a given trace is feasible. For example, we can check whether a particular trace for the successful establishment of the mutual authentication between Alice and Bob is feasible.

During the manual exploration or automatic check, the animated protocol often deadlocks (that is, no further enabled events). This happens because we model only one protocol run (instead of an infinite number of sessions) for both Alice and Bob. If, for example, a message faked by the intruder interferes with the standard protocol run for Alice or Bob, they will wait for the next message, which may never be possible. This modelling of one protocol run using a single initiator role for Alice and a single responder role for Bob will not affect whether an attack exists and could be found according to Lowe~\cite{Lowe1996}. He proved that if there is an attack in the NSPK protocol, the attack must be in this simple version. So, if NSPK3 is secure, then the general NSPK is also secure. 

\paragraph{NSLPK3.} In Lowe's corrected protocol~\cite{Lowe1996}, B's identify is additionally appended to Message 6 in \Cref{fig:nspkp} to stop the intruder from replaying this message from B to A because A expects this message from the intruder (with whom A is told to establish the authentication in the attack) but it is from B. We 
changed the message sent in \cspcode{Responder}\isaref{{\codereponspk\#L341}} to $\aenc{\langle \cspcode{MNon na},{\cspcode{MMon nb}}, \cspcode{MAg B}\rangle}{\cspcode{PK A}}$) and our animation could not find such an attack.

\paragraph{NSPK7.}
For the original NSPK7 shown in \Cref{fig:nspkp}, Alice and Bob do not know each other's public key. They retrieve the public keys from a server. So a new agent \cspcode{Server} is added to \cspcode{dagent}. The server uses its private key to sign messages and sends them to Alice and Bob, who know the server's public key. So we add a new type of message, \cspcode{MSig}, in \cspcode{dmsg}, denoted as $\digsig{m}{k}$.
\newcommand{\dagentnspkseven}{\isalink{\codereponspkmsgseven\#L11}}
\newcommand{\dmsgnspkseven}{\isalink{\codereponspkmsgseven\#L168}}
\begin{alltt} 
\isakwmaj{datatype} dagent = Alice | Bob | Intruder | Server \dagentnspkseven
\isakwmaj{datatype} dmsg = ... | MSig (msd:dmsg) (msk:dskey) \dmsgnspkseven
\end{alltt}
We also assume Alice and Bob use a private network to communicate with the server so the intruder cannot intercept the messages. For this purpose, we add two secure channels \cspcode{send\_s} and \cspcode{recv\_s} in \cspcode{Chan}\isaref{\codereponspkmsgseven\#L370}. The intruder does not know Alice and Bob's public keys and uses these secure channels to get them.
Our animation shows that NSPK7 is vulnerable to the same man-in-the-middle attack. 

\section{Diffie–Hellman Key Exchange Protocol}
\label{sec:dhkep}
The Diffie-Hellman (DH)~\cite{Diffie1976} protocol aims to establish a shared secret between two agents using agreed information over an insecure network. Then they can use the secret to encrypt further messages or derive encryption keys. 
DH is based on modular exponentiation $g^a \mod p$ where $g$ is the base, $a$ is the power or exponent, and $p$ is the modulus. Both parties publicly (that is, the intruder also knows) agree to use $g$ and $p$. The $a$ is private and only known to the agent who owns it.

We illustrate the original protocol (DH) in \Cref{fig:dh_original}, which is vulnerable to the man-in-the-middle attack, and a variant (DHDS) of the protocol in \Cref{fig:dh_sign} that uses the digital signature to tackle the attack where the modulus $p$ part is omitted. The general procedure of a DH protocol run includes 
\begin{enumerate*}[label=(\arabic*),]
    \item both $A$ and $B$ send $g^{na}$ and $g^{nb}$ (where $na$ and $nb$ are their private nonces) to each other, 
    \item they compute $k_A = g^{nb^{na}}$ and $k_B = g^{na^{nb}}$, and so $k_A = k_B$ (that is, they get a shared secret), and
    \item $A$ sends a secret (symmetrically) encrypted by $k_A$, and then $B$ can use the same $k_B$ to decrypt it.
\end{enumerate*}
In DHDS, A knows B's public key and expects a message from B that is signed using B's private key. The intruder cannot forge B's digital signature without knowing B's private key.

\begin{figure}[t]
    \subfloat[Original DH protocol]{
        \label{fig:dh_original}
  \begin{minipage}[b]{0.41\linewidth}
\begin{center}
    \begin{enumerate}[label={{{\arabic*}.}}]
        \item {$ A \to B: g^{na} $}
        \item $B$ computes $k_B = g^{na^{nb}}$
        \item $ B \to A: g^{nb}$
        \item $A$ computes $k_A = g^{nb^{na}}$
        \item $ A \to B: \enc{s}{k_A} $
        \item $B$ decrypts it using $k_B$ to get $s$ 
\end{enumerate}
\vspace*{23pt}
\end{center}
  \end{minipage}
}
\subfloat[DH with digital signature]{
        \label{fig:dh_sign}
  \begin{minipage}[b]{0.53\linewidth}
\begin{center}
    \begin{enumerate}[label={{{\arabic*}.}}]
        \item {$ A \to B: \mcomp{\digsig{g^{na}}{\sk{k}{A}}}{\pk{k}{A}} $}
        \item $B$ uses received $\pk{k}{A}$ to verify the message and gets $g^{na}$; $B$ computes $k_B = g^{na^{nb}}$
        \item $ B \to A: \digsig{g^{nb}}{\sk{k}{B}}$
        \item $A$ initially knows $\pk{k}{B}$ and verifies the message and gets $g^{nb}$; $A$ computes $k_A = g^{nb^{na}}$
        \item $ A \to B: \enc{s}{k_A} $
        \item $B$ decrypts it using $k_B$ to get $s$ 
\end{enumerate}
\end{center}
  \end{minipage}
}
    \caption{Diffie-Hellman key exchange protocol}
    \label{fig:dh}
\end{figure}

In Isabelle, we add three types of messages in \cspcode{dmsg}: \cspcode{MSEnc} for symmetric encryption, \cspcode{MExpg} for the base $g$, and \cspcode{MModExp} for modular exponentiation with corresponding annotations: $\enc{m}{k}$, $g$, and $(g_m)^b$ 
\newcommand{\dmsgdh}{\isalink{\coderepodhmsg\#L191}}
\begin{alltt} 
\isakwmaj{datatype} dmsg = ... | MSEnc (msem:dmsg) (msek:dmsg) \(\dmsgdh\)
    | MExpg | MModExp (mmem:dmsg) (mmee:dnonce) 
\end{alltt}

For the implementation (in \cspcode{breakm} \isaref{\coderepodhmsg\#L437}) of the decryption break-down rule in \Cref{table:inference}, an encrypted message $\enc{m}{k}$, when the key $k$ is a \cspcode{MModExp} like $g^{{na}^{nb}}$, can be decrypted only if $g^{na}$ and $nb$ are known, or $g^{nb}$ and $na$ are known, or $g$, $na$, and $nb$ are all known.

For the build-up rule, we implement functions \cspcode{mod\_exp1}\isaref{\coderepodhmsg\#530} and \cspcode{mod\_exp2}\isaref{\coderepodhmsg\#520} to compose messages (from a list of nonces) that are resulted by applying modular exponentiation up to once and twice.

\begin{figure}[tb]
\begin{lstlisting}[language=Animation, caption={}, label=]
/*AReach 15,  %Leak PK Alice%
Reachability by Auto: 15
  Events for reachability check: ["Leak PK Alice"]
  Events for monitoring: []
*/*** These events ["Terminate"] are reached! ***
Trace: [Send [Alice=>Intruder] g^N Alice, Send [Bob=>Intruder] g^N Bob, Recv [Alice<=Intruder] g^N Intruder, Recv [Bob<=Intruder] g^N Intruder, Send [Alice=>Intruder] {PK Alice}^S_g^N Intruder^N Alice, Leak PK Alice, Recv [Bob<=Intruder] {PK Alice}^S_g^N Bob^N Intruder, Terminate, ]
\end{lstlisting}
\caption{\label{fig:animate-dh1-violet-secrecy} One trace of the violation of secrecy in the original DH.}
\end{figure}

We use the animation to find an attack in the original DH by automatic reachability check of the \cspcode{terminate} event. One counterexample (among several) is shown in \Cref{fig:animate-dh1-violet-secrecy} where the encrypted secret (\cspcode{PK Alice}) is leaked, but Bob believes he gets the encrypted message from Alice. The trace shows that the intruder establishes a shared key ($g^{{ni}^{na}}$) with Alice and a shared key ($g^{{ni}^{nb}}$) with Bob where $ni$, $na$, and $nb$ denote \cspcode{N Intruder}, \cspcode{N Alice}, and \cspcode{N Bob}. The secret (\cspcode{PK Alice}) sent by Alice is decrypted by the Intruder (using the shared key with Alice), and then the Intruder re-encrypts the secret using the shared key with Bob. Bob finally decrypts it and believes it is from Alice. This is the man-in-the-middle attack.

We also used the animation to check DHDS and found only two traces of the successful termination, neither of which would leak the secret. This concludes that DHDS is secure for the attack.

\section{Conclusion and future work}
\label{sec:concl}
In this paper, we presented a novel approach to automatically generate sound animation in the Dolev-Yao attack model, arguably the most adopted model for symbolic analysis for security protocols.
We implement our analysis in ITree-based CSP in Isabelle, to support verification guided by users.
Our animator is a verified model checker, able to check reachability using automatic exhaustive or random search. Additionally, it supports interactive exploration and feasibility checks, which are usually not provided by model checkers.

We showcased our approach with two traditional case studies, the Needham-Schroeder protocol and the Diffie-Hellman key exchange protocol.
We model them in five variants (in total) and verify such variants via the generated animators.
Once the animator is generated, it can be used by users who do not need to know the formal language (Isabelle).
Our animator can find the attacks in the flawed variants and guarantee these attacks do not exist in the revised variants.

Improvements to our current approach can take several directions.
One future work is to develop a general framework to combine different features implemented in the examples, such as asymmetric and symmetric encryption, digital signature, and modular exponentiation. We will also introduce a hash function.
Another improvement is to support a multiple (or possibly infinite) number of sessions, as our model for each agent supports only one.
This might be done by implementing a CSP operator similar to the replicate operator from the $pi$-calculus (e.g., as implemented in ProVerif).
Also, our current approach supports the workflow illustrated in \Cref{fig:workflow}, where formal researchers are still required to model security protocols.
We can create a domain-specific language (DSL) for the modelling of security protocols, then automate the transformation from the DSL to ITree-based CSP.
The ambitious and challenging aim here is to fully automatize the workflow and allow protocol designers to carry out user-guided verification without {\em any} involvement of formal verification experts.

Our work has many potential applications.
Our future interests include the verification of security protocols used in 5G or 6G networks, particularly in the areas with new architectures, such as Open-RAN, or with resource-constrained devices typical, for example, in edge computing.
Security analysis in such may require flexible attack models rather than the Dolev-Yao model.

Finally, animation has certain limitations in terms of enumerable and finite data types such as agents, keys, nonces, and messages, as well as intensive space and computation time requirements to infer messages for the intruder, which is mitigated by imposing the number of times a build-up rule can be applied in \cspcode{build\_n}. These are due to the executable nature of animation. They, however, are not the limitations of ITrees and ITree-based CSP. We can use general {\em datatypes} or {\em codatatypes} 
and inductive sets in Isabelle/HOL to define (infinite) inferred messages.
Such a model can be verified using theorem proving based on the operational and denotational semantics of ITree-based CSP~\cite{Foster2021}. This, however, creates a hole between models for theorem proving and for animation, which can be filled by using functional algorithm and data refinement in Isabelle's code generation (thanks to its equational logic). This is part of our future work to integrate theorem proving and animation together.

\noindent \textbf{Acknowledgements.}  

The EPSRC and DSIT support this work through the Communications Hub for Empowering Distributed Cloud Computing Applications and Research (CHEDDAR) under grants EP/X040518/1 and EP/Y037421/1.

  We thank Jim Woodcock and Simon Foster for various discussions of this work and helpful comments. 

\newpage
%

\bibliographystyle{splncs}
\bibliography{main.bib}


\ifdefined \CHANGES \indexprologue{%
  This index lists for each comment the pages where the text has been modified to address the comment. Since the same page may contain multiple changes, the page number contains the index of the change in superscript to identify different changes. Finally, the page number contains a hyperlink that takes the reader to the corresponding change.%
}%
\printindex[changes] \fi

\end{document}